# Title: Real-space investigation of polarons in hematite $Fe_2O_3$

Jesus Redondo,[1,2,3] Michele Reticcioli,[4] Vit Gabriel,[1] Dominik Wrana,[1] Florian Ellinger,[4] Michele Riva,[2] Giada Franceschi,[2] Erik Rheinfrank,[2] Igor Sokolovic,[2] Zdenek Jakub,[2] Florian Kraushofer,[2] Aji Alexander,[1] Laerte L. Patera,[5,6] Jascha Repp,[5] Michael Schmid,[2] Ulrike Diebold,[2] Gareth S. Parkinson,[2] Cesare Franchini,[4,7] ** Pavel Kocan,[1] Martin Setvin[1,2] *

[1] Department of Surface and Plasma Science, Faculty of Mathematics and Physics, Charles University, 180 00 Prague, Czech Republic
[2] Institute of Applied Physics, TU Wien, 1040 Vienna, Austria
[3] Institute of Physics, Czech Academy of Sciences, Cukrovarnická 10, Prague 6, 162 00, Czech Republic
[4] University of Vienna, Faculty of Physics, Center for Computational Materials Science, Vienna, Austria
[5] Institute of Experimental and Applied Physics, University of Regensburg, 93040 Regensburg, Germany
[6] Institute of Physical Chemistry, University of Innsbruck, 6020 Innsbruck, Austria
[7] Dipartimento di Fisica e Astronomia, Università di Bologna, 40127 Bologna, Italy.

\* [martin.setvin@mff.cuni.cz](martin.setvin@mff.cuni.cz)
\*\* [cesare.franchini@univie.ac.at](cesare.franchini@univie.ac.at)

**In polarizable materials, electronic charge carriers interact with the surrounding ions, leading to quasiparticle behaviour. [1,2] The resulting polarons play a central role in many materials properties including electrical transport, [3-5] optical properties, [6] surface reactivity[7,8] and magnetoresistance, [9,10] and polaron properties are typically investigated indirectly through such macroscopic characteristics. Here, noncontact atomic force microscopy (nc-AFM) [11] is used to directly image polarons in $Fe_2O_3$ at the single quasiparticle limit. A combination of Kelvin probe force microscopy (KPFM) [12] and kinetic Monte Carlo (KMC) simulations shows that Ti doping dramatically enhances the mobility of electron polarons, and density functional theory (DFT) calculations indicate that a metallic transition state is responsible for the enhancement. In contrast, hole polarons are significantly less mobile and their hopping is hampered further by the introduction of trapping centres.**

Hematite has long been touted as a promising photoanode for photocatalytic water splitting due to its ≈ 2 eV band gap, but its practical use is hindered by sluggish reaction kinetics. [13-15] This is partly because photogenerated electrons and holes interact with the crystal lattice forming localized quasiparticles known as polarons, which can only diffuse given an activation energy. The performance is often enhanced by doping, which increases the polaron concentration and, seemingly, their mobility. [15-17] This is counterintuitive, however, because defects usually impede electron transport in materials. Nevertheless, similar observations have been reported for other materials including CuO:Li, [18] $BiVO_4$, [19] and NiO. [20] The common property of all these systems is their antiferromagnetic spin ordering combined with nonmagnetic dopants. [15-17]

The experiments described here were performed on the (1-102) surface of $Fe_2O_3$, a nonpolar face that preferably terminates with the surface in a simple (1×1) configuration.[21,22] Thin films (50–100 nm) of a purposely-doped hematite were grown[23] on natural $Fe_2O_3$ single crystals by pulsed laser deposition, see the sketch in Fig. 1A. Electron and hole polarons have been studied on *n*-doped (Ti doping with concentrations 0, 0.03, 0.7 and 3 atomic percent) and *p*-doped (0.1 atomic % Ni) hematite, respectively.[24,25]

An example of the surface topography is shown in Fig. 1D, with more details discussed in Fig. S1. $Fe_2O_3$ barely conducts at room temperature and becomes insulating at cryogenic temperatures when diffusion of charge carriers (polarons) is frozen out in the lattice. Consequently, the material can be locally charged with an AFM tip, as illustrated in Fig. 1B. Excess electrons or holes can be injected from the tip when a positive or negative bias is applied to the sample, respectively, while the tip is in tunnelling contact (≲1 nm distance). Kelvin parabolas (where maxima correspond to the local contact potential difference, LCPD) yield the electrostatic potential above the surface,[12] provided that the tip-sample distance is large enough to prevent electron exchange by tunnelling (≳3 nm). The black and red parabolas demonstrate how the LCPD changed after injecting ~80 holes into the surface by tunnelling with a negative sample bias. The resulting hole-polarons are stable for many hours, see the orange parabola in Fig. 1B. The small vertical shift is attributed to thermal drift. A characteristic feature of polarons is that their mobility can be promoted by light (unlike for structural defects). This is demonstrated by the blue parabola, where the LCPD approaches its initial value upon illuminating the sample by visible light (regular room illumination; note that hematite has a bandgap within the visible region (1.9 – 2.4 eV)).[14,22]

The process of charge injection can be studied with single-particle precision, as illustrated in Figure 1C on hole polarons. The tip was positioned ~0.5 nm from the surface and a bias voltage of -1.2 V was applied to the sample, resulting in the stepwise formation of hole polarons. The frequency shift shows stepwise increases. It is established that such steps correspond to single-electron charging events.[26] Here we attribute it to the tunnelling of single holes into the region below the tip. All the newly formed polarons appear to be created very close to the tip apex because all the steps have a comparable magnitude. As more and more polarons are injected, they cause a Coulomb blockade[27] that suppresses further tunnelling: The rate of the injection events decreases with the number of polarons already formed. This is illustrated by plotting different sections of the injection curve in Fig. 1C. The cloud of polarons relaxes during the injection process, as evidenced by the apparent drift of the measured frequency shift in between the injection events. These relaxations are promoted by the electric field of the tip and by Coulomb repulsions among polarons. More details and analogous curves for electron injection are provided in Fig. S2.

Fig. 1E shows that single hole polarons can be placed at specific positions. Different numbers of holes (1 to 5) have been injected in locations marked by the respective numbers of dots. The image in Fig. 1E is a difference of constant height Δ*f* (frequency shift) images measured at identical parameters before and after the polaron injection; full details of this experiment are given in Figs. S3 and S4. Areas with increasing brightness form for one to three polarons. For higher numbers, the quasiparticles start to spread and scatter due to repulsive Coulomb interactions. Note that the apparent spot size of ~5 nm is attributed to the effective size of the tip apex; the calculated polaron size is less than 1 nm. The calculated charge distributions of one hole and one electron small polaron are shown in Fig. 1F. The hole is localized mostly on a single Fe atom (33%), strongly hybridizing with the surrounding oxygen atoms.[28] For the electron polaron, the projection of the polaronic states into the local atomic orbitals shows that two Fe atoms equally share 80% of the electron charge (the two Fe atoms show the typical $d^5$ occupation of the spin majority channel, and the polaronic electron in the minority spin channel, see sketch in Fig. S5).

The key challenge in polaron physics is understanding their complex kinetics. It is a source of many fascinating polaron-related phenomena, but can also hinder the performance of materials in applications, which is specifically the case for hematite and photocatalysis.[29] To study polaron kinetics, we have focused on statistical sets of many (~hundreds) charge carriers injected into the surface. This approach allows us to neglect the influence of tip-induced polaron manipulations. The spatial outreach of such injected electron- and hole-polaron clouds is shown in the LCPD maps in Figs. 2A and 3A, respectively. Note that these images are micrometers wide. The cloud of holes (Fig. 3A) is more confined in space compared to electrons (Fig. 2A), which is attributed to a significantly lower mobility of hole polarons.[28]

Figs. 2 and 3 summarize the experimental and simulation results on electron- and hole-polaron kinetics, respectively. In Fig. 2, each experiment began by injecting ~300 electrons and measuring an LCPD map. Subsequently, the sample was gradually annealed to increasing temperatures. In each annealing cycle, the temperature was kept at the given value for 10 minutes and decreased back to 4.7 K for imaging. The cloud evolves when the polaron mobility is enhanced by thermal excitations. Line profiles across the cloud are shown in Fig. 2C. The clouds always maintain circular symmetry. The temperature dependence of the cloud profile shows that the dominant effects responsible for polaron migration are thermal excitations combined with electrostatic interactions among the polarons (the role of electrostatics is linked to the wide temperature range of the whole process). The stability of such polaron clouds depends critically on doping: Repeating the same experiment for several doping levels of Ti (see Figs. 2C, 2E and S6) shows a dramatic dependence. For a 3% doping level, the cloud scatters completely at temperatures above 10 K, while undoped samples require ~40 K for an analogous effect, highlighting the key role of Ti in enhancing the electron polaron mobility.

Kinetic Monte Carlo (KMC) simulations were performed to extract the parameters of electron-polaron migration (Fig. 2B,D,F). The model (detailed description in methods) considered electrostatic interactions among all polarons in the slab, Arrhenius hopping to the nearest-neighbour site with an activation energy $E_A$ and a frequency prefactor $v_0$, and considered the interface between the hematite (dielectric permittivity of $\varepsilon_r = 20$)[13] and vacuum. For electron polarons, the KMC-simulated electrostatic potentials as a function of the annealing temperature (Fig. 2D,F) show an excellent match with the experiment. An example of evolution of the simulated polaronic cloud is shown in Fig. 2B and in the supplementary movie SM1.

Only two parameters were fitted in the simulation, the activation energy $E_A$ and the preexponential factor $v_0$. The best fit was obtained for a prefactor $v_0 = 10^{6\pm1}$ s$^{-1}$, much lower than the typical value of $10^{13}$ s$^{-1}$.[23] The simulation results for $v_0 = 10^{11}$ s$^{-1}$ are shown in Fig. S7. Table 1 shows the estimated activation energies as a function of the Ti doping level. The $E_A$ is shown for $v_0 = 10^6$ s$^{-1}$ and $v_0 = 10^{11}$ s$^{-1}$, *i.e.,* cases where $v_0$ was fitted or fixed, respectively.

The used KMC models examined the cases of purely surface diffusion, purely bulk diffusion, and diffusion limited to the Ti-doped layer (see the sample geometry in Fig. 1A; the Ti-doped layer has a higher polaron mobility than the underlying undoped sample). A good match with the experimental data was obtained for diffusion restricted the doped layer or for pure bulk diffusion. Pure surface diffusion can be excluded. The role of surface/bulk diffusion is discussed in detail in Fig. S8.

| | undoped | 0.03 % Ti | 0.7% Ti | 3% Ti | 0.1% Ni (holes) |
|---|---|---|---|---|---|
| $E_A$ for $v_0=10^{11}$ s$^{-1}$ | 70 meV | 50 meV | 24 meV | 17 meV | 200 meV |
| $E_A$ for $v_0=10^6$ s$^{-1}$ | 43 meV | 30 meV | 15 meV | 11 meV | |

*Table 1: Activation energies for electron polaron hopping in nominally undoped and Ti-doped material, extracted from the experimental data using KMC simulations. The last column shows results for hole hopping in a Ni-doped sample.*

The migration of hole-polarons is summarized in Fig. 3, measured on a film doped with 0.1% Ni. (Note that hole-polarons could be injected into Ti-doped films as well but their kinetics cannot be studied when the more mobile electron-polarons dominate and annihilate the holes). Fig. 3B shows experiments performed at a base temperature of 4.7 K. Here the electrostatic potential measured above the surface did not undergo a substantial decay. Figure 3C shows an experiment performed at a base temperature of LN$_2$ (77 K) and with a sample annealed up to 140 K. Initially there is a slight decay of the electrostatic potential, but the cloud does not evolve further after annealing to higher temperatures. This behaviour has been observed several times. It is noteworthy that the hole-polaron clouds could not be fully erased by light illumination, which was always possible for electron-polaron clouds.

KMC simulations show that the diffusion of hole polarons cannot be described by a single activation energy. The simulations considered various concentrations of trapping centres, *i.e.*, lattice sites that the polaron cannot escape at the highest experimental temperature of 140 K. A trap concentration of 10$^{-3}$ provides a good match to the experimental data; different trap concentrations are shown in Fig. S9. The activation energy for a hole-polaron estimated from the initial decay of the cloud is 0.2 eV, assuming a pre-factor of 10$^{11}$ s$^{-1}$. This matches the activation energy for hole-polaron diffusion estimated by DFT. [28,30]

The concept of hole-traps is well established in photocatalysis. One can speculate about the origin of such trapping centres: (i) They can be defects present in the lattice before the polaron injection. It is noteworthy that the calculated concentration of 10$^{-3}$ matches the level of Ni doping. (ii) The holes can weaken the lattice oxygen bonds and promote defect formation. iii) The experiments were performed on a p-doped layer grown on a bulk crystal, which is presumably slightly *n*-doped due to oxygen vacancies. The initial decay of the hole-cloud could be due to the hole annihilation by underlying electrons, which stops later due to the formation of a *p-n* junction.

For electron polarons, an intriguing relation between the activation energy $E_A$ and doping by trace Ti impurities was observed. The $E_A$ decreases by a factor of four as the Ti doping increases from 0 to 3% (3% is approximately the optimum doping for Fe$_2$O$_3$:Ti). [31] Such *defect-enhanced* polaron mobility is unusual, but has been observed in materials with antiferromagnetic ordering. [18-20] Our data indicate that the effect of dopants is nonlocal: Doping levels below ~0.1% Ti already influence the electron mobility significantly; this corresponds to one Ti dopant in a grid of 10×10×10 Fe atoms. This can be hardly explained by a local modification of the activation energies near the dopants, as considered previously.[28,32-34] Instead, based on our DFT calculations, we propose that the mobility enhancement is promoted by a delocalized (free-electron-like) transition state, see Fig. 4. There are two types of hops: The easier path occurs within one ferromagnetic plane (Fig. 4A,C,E) through spin-alike Fe sites, where the transition state is polaronic. In this case, the Ti doping slightly decreases the hopping energy barrier (Fig. 4A). The more difficult (rate-limiting) step leads across the antiferromagnetically ordered planes (Fig. 4B,D,F), since the polaron passes through Fe atoms with no available states (in the antiferromagnetically ordered plane, all 3d states with the same spin as the polaron are occupied). Here the transition state crosses the Fermi level (see the metallic polaronic peak in Fig. 4D): the polaron becomes destabilized at energies smaller than the hopping energy barrier, thus the electron acquires a metallic character interrupting the hopping across the antiferromagnetic plane. The delocalized electron might relocalize in any other available Fe site with local distortions suitable to trap the charge carrier. This process involves the momentarily destabilization of a polaron; the diffusion via a metallic state is a process known as random flight.[35,36] With such a free-electron-like transition state, ballistic movement of the electron could occur in direction of the electric field, which would be consistent with the isotropic shape of the polaronic clouds observed experimentally. The role of the Ti dopant is to provide the initial and final polaronic states with a good overlap of their electronic wavefunctions that enhances the probability to activate this process (see Fig. S10 in the SI).

The calculated values of activation energies (Fig. 4 A,B) are in-line with the experimental data in Table 1 for the undoped material. The presence of Ti leads to a certain decrease of the barrier, yet the experimentally observed effect is quantitatively stronger. This is attributed to inability to find the optimal transition state. Experiments indicate that the frequency prefactors are below the standard value of $10^{13}$ s$^{-1}$ (derived from the frequencies of lattice vibrations) and are possibly as low as ~$10^6$ s$^{-1}$. The low prefactors might partially originate from the entropy of a transition state that involves a concerted motion of many atoms, thus is unlikely in the transition state theory.[37] However, the value of the prefactor should be taken with caution because it could also originate from an uneven spatial distribution of activation energies caused by dopants and impurities, or also from the use of a simplified model for the electrostatic polaron-polaron repulsions in the KMC simulations, see Fig. S11.

In summary, this work used noncontact AFM to image both electron- and hole-polarons in real space, down to the single quasiparticle limit. AFM represents a novel tool for non-destructive probing of polaron properties at the atomic scale, investigating their kinetics induced by electric fields, light and thermal excitations, as well as tackling details about the transport mechanisms and charge trapping. This was demonstrated here by identifying the random flight mechanism as an important type of electron-polaron transport in hematite. The methodology introduced here may open a new way to developing single-electron devices, multistate memories for neuromorphic computing,[38] as well as understanding fundamental mechanisms in catalysis.

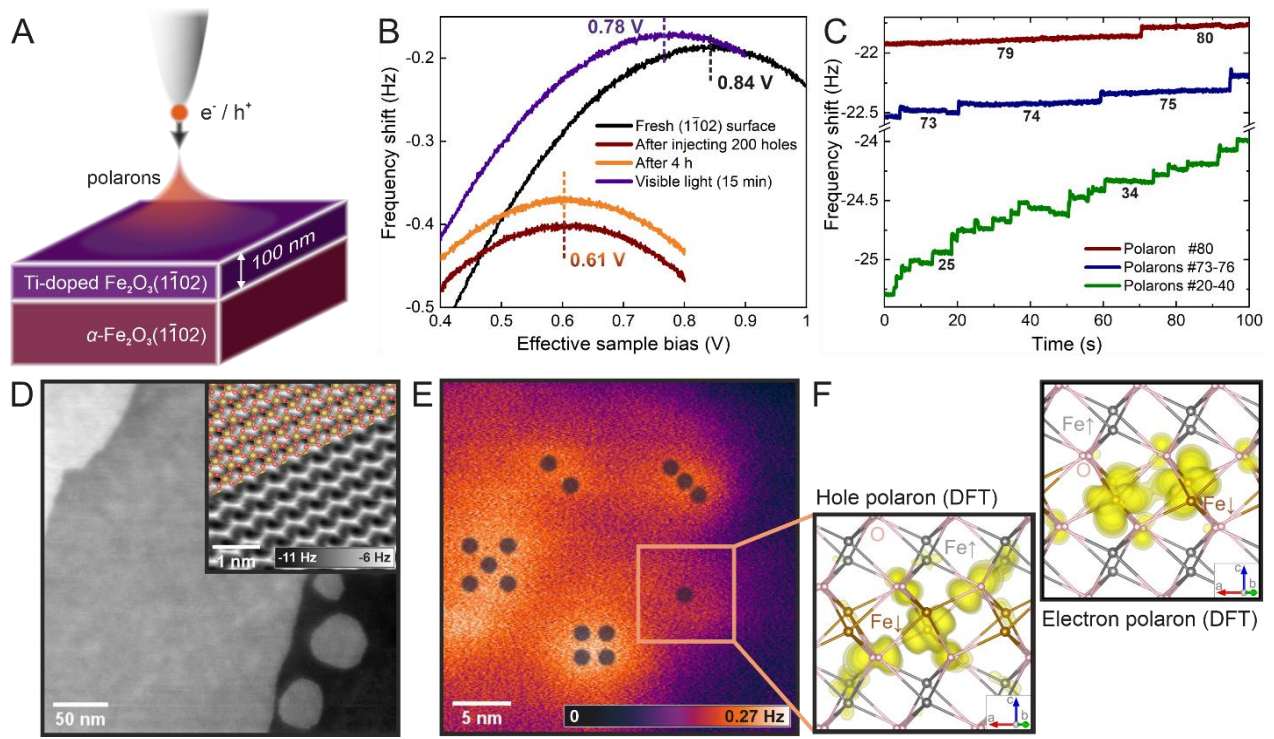

*Figure 1 Charge injection in hematite.* A) Sketch of the experimental setup. B) Kelvin parabolas showing the LCPD of the pristine surface (black), after injecting ~80 holes (red) into a 0.1% Ni-doped surface, after waiting for 4 hours (orange), and after illuminating the surface by visible light (blue). C) Time-dependence of the frequency shift when injecting holes into the surface at effective $V_S = -1.2$ V; each jump indicates formation of one individual hole-polaron. D) Topographic nc-AFM image of the surface obtained at a constant detuning of -0.57 Hz, $A$ = 600 pm, $V_S$ = 80 mV. The inset shows an atomically resolved constant-height nc-AFM image, overlaid with the structural model. E) Map of electrostatic forces induced by various numbers of hole-polarons injected into the surface at locations marked by the respective numbers of dots. F) Calculated charge densities of a single hole-polaron and an electron-polaron in bulk hematite. All experimental data were measured at $T$ = 4.7 K.

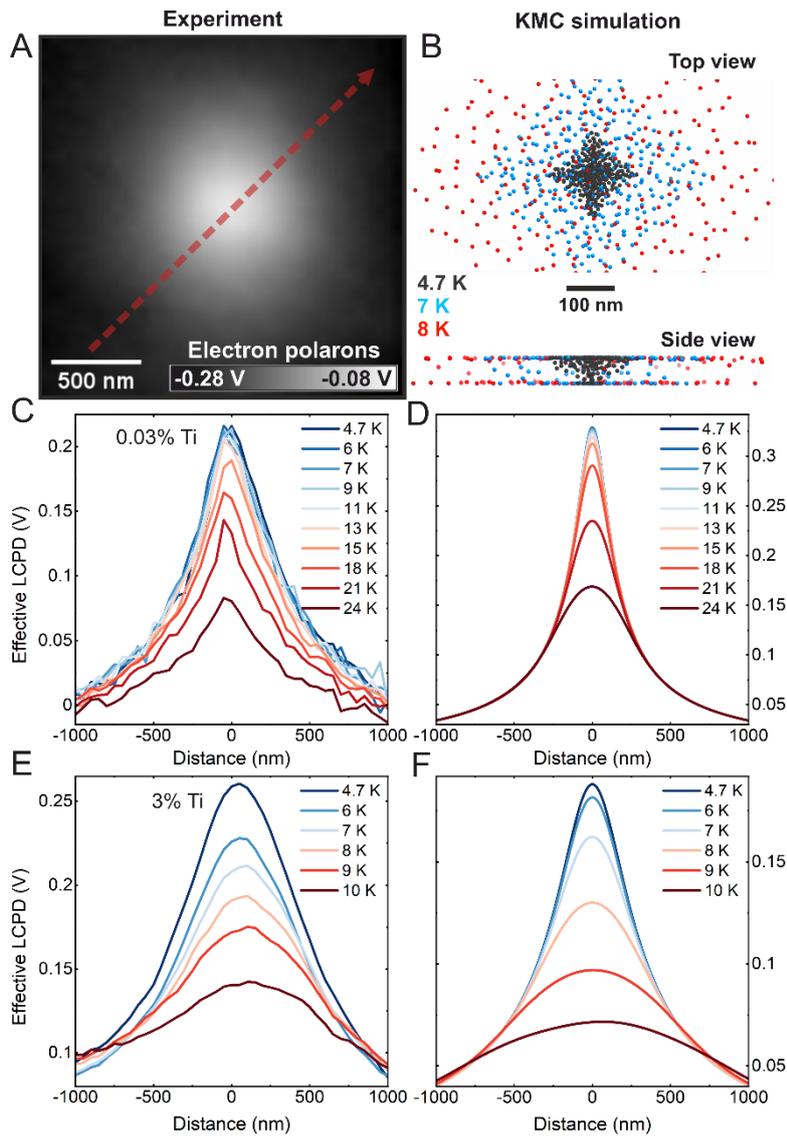

***Figure 2: Thermally activated diffusion of electron-polarons.*** *A) Map of LCPD induced by ~300 electron-polarons injected into the hematite surface doped with 3% Ti. B) Evolution of the electron-polaron cloud during gradual annealing, as simulated by KMC (mimicking the experimental procedure of annealing for 10 minutes at each temperature). C,E) effective LCPD line profiles measured across a cloud of ~300 electron polarons, measured after gradual annealing to increasing temperatures. Ti doping levels were 0.03% and 3% Ti, respectively. The path of the line profile is denoted by the dashed red arrow in panel A. D,F) KMC simulations corresponding to panels C and E, respectively. The best agreement between the KMC simulations and the experimentally obtained LCPD profiles was obtained using the activation energies listed in Table 1. The dopants effectively decrease the activation energy for hopping.*

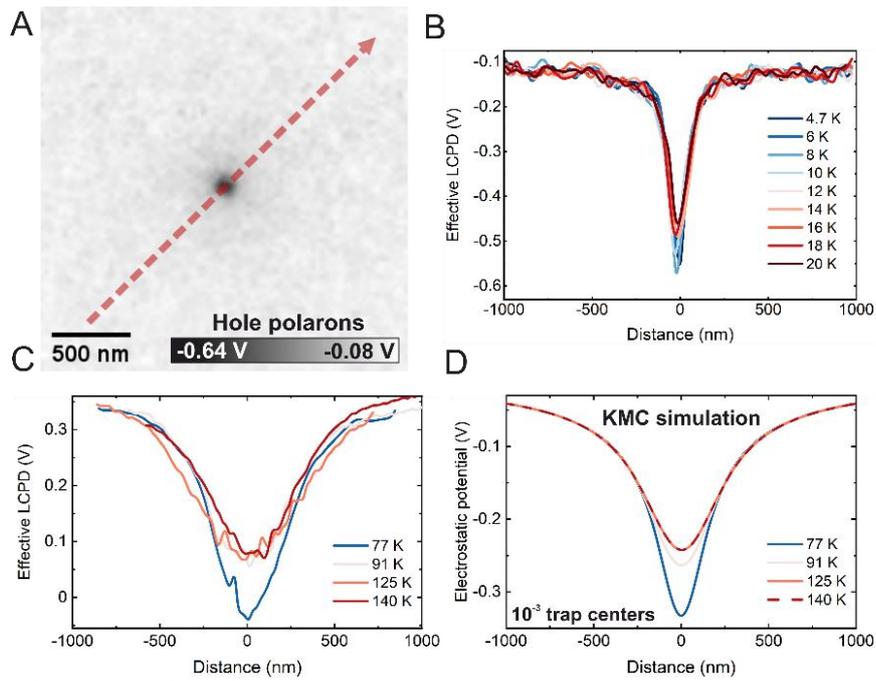

*Figure 3: **Diffusion of hole polarons,** measured in a p-doped film, 0.1% Ni. A) Map of LCPD induced by ~230 hole polarons injected into the surface at T = 4.7 K. B,C) effective LCPD from a cloud of ~300 and 150 hole polarons, respectively, and its evolution upon annealing up to 20 and 140 K, respectively. The clouds were injected at T = 4.7 K and 77 K, respectively. D) KMC simulation of the hole diffusion denoted in panel C, with introduction of hole-trapping centres with a concentration of $10^{-3}$. For other trap concentrations see Fig. S8.*

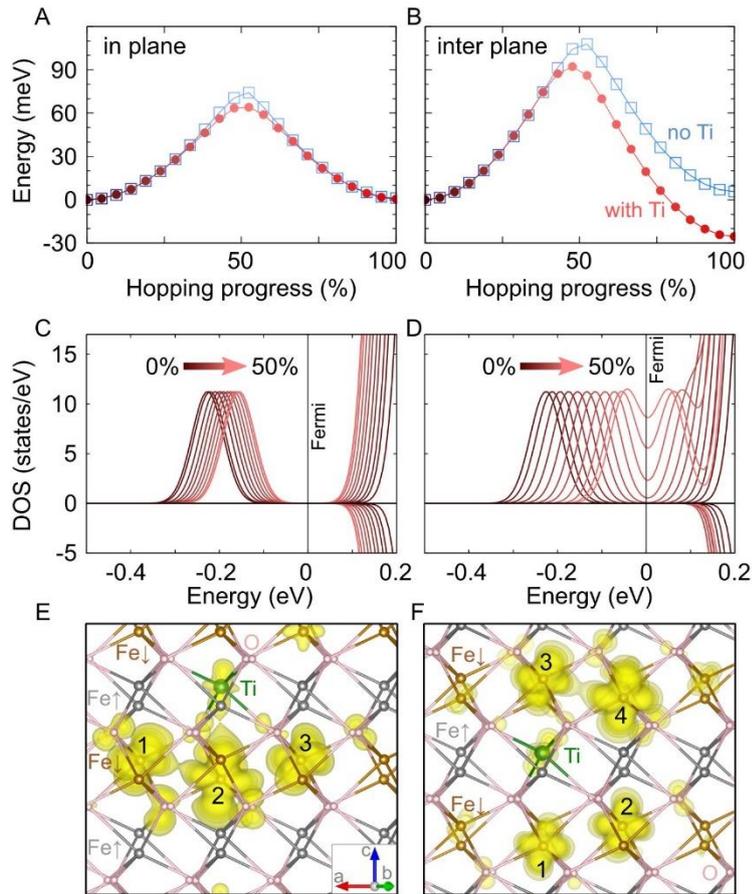

*Figure 4: Electron-polaron hopping in hybrid-functional DFT. Panels A,C,E and B,D,F relate to the in-plane and inter-plane hopping, respectively. Panels A,B show the energy barriers as a function of the progressing lattice distortion without (blue) and with (red) Ti dopants. Panels C,D show the spin-resolved DOS during the hopping process (from 0% to 50% in gradient of red); only the case with the Ti dopant is shown. Panels E,F show the isodensity of the transition state (i.e., the polaron at 50% progress); O atoms are shown in pink, Ti in green, Fe in brown and grey to distinguish between opposite orientation of the local magnetic moments. In E, the polaron is hopping in-plane from the initial configuration at sites 1, 2 (0%) towards the final state at sites 2,3 (100%). F corresponds to an interplanar hop from sites 1,2 (0%) towards 3,4 (100%).*

**Methods**

The samples were grown ex-situ by pulsed laser deposition;[23,39] details are described in the SI. The films of Ti-doped hematite were 70 to 100 nm thick and the doping levels are estimated as (0.035±0.002), (0.77±0.06), and (3.09±0.24) atomic % Ti. The Ni doped film was 100 nm thick, and the Ni doping concentration was (0.1±0.01) at.%. In the text, all doping concentrations are rounded down.

Prior to the AFM measurements, surfaces were cleaned by cycles of sputtering and annealing to 640 °C in a partial oxygen pressure of $5 \times 10^{-4}$ Pa.[40] A uniform (1×1) surface termination was obtained on the Ti-doped samples. The Ni-doped samples typically had residues of the (2×1) reconstruction.[22] Notably, all results related to polaron kinetics appear insensitive to the surface quality and surface termination. This supports the picture of bulk migration deduced from the KMC simulations.

The combined STM/AFM measurements were performed on commercial cryogenic systems from ScientaOmicron, either the Polar or the LTSTM, in ultrahigh vacuum (UHV). A differential preamplifier[11] for the deflection signal was used. Defined tip conditions were obtained by treatment on a Cu (110) surface. A base temperature of 4.7 K was used for most experiments, except for the investigations of hole-polaron mobility (77 K). LCPD maps were obtained by measuring a grid of Kelvin parabolas and fitting the data. The range of applied biases was strictly controlled to avoid the manipulation of the polarons by the tip electric field. Controlled sample annealing inside the STM/AFM head was done by a heater placed behind the sample. After ramping the temperature to the setpoint, the sample was kept at the desired temperature for 10 minutes and cooled back to the base temperature. All experiments were performed in darkness by covering all windows on the UHV chamber, switching off the ion gauge and keeping the room dark; visible light promotes polaron redistribution. Whether this effect is promoted by exciting the existing polarons by light, or by generating new electron-hole pairs that interact with the existing polarons, was not further pursued.

For bulk-insulator samples the actual potential drop between a probe tip and the sample is difficult to ascertain, since the field penetrates into the insulating sample. We have estimated that approximately 20% of the applied bias is in the junction and 80% is inside the sample. Several methods were used for this estimation, see the SI. Importantly, all voltages and LCPD values mentioned in the text are already corrected for this effect, *i.e.*, they are equal to 1/5 of the bias applied on the sample plate.

The collective behavior of the polarons injected into the surface was simulated by a simplified KMC model. The electron and hole polarons were approximated as point charges equivalent to +1 or -1 $e$, respectively. The hematite structure is simplified to a cubic network of sites with a lattice parameter of 12 Å (the size of the grid is discussed in detail in the SI, Fig. S11), in which the polarons may be located, at most one polaron per site. The polarons are allowed to hop to unoccupied neighboring sites with the rate $R = v_0 \exp[-(E_A+\Delta E/2)/k_B T]$, where $v_0$ is the frequency prefactor, $E_A$ is the diffusion barrier of the isolated (non-interacting) polaron, fitted separately for each hematite doping level, $k_B$ is the Boltzmann's constant and $T$ is the temperature. $\Delta E$ is the difference of the electrostatic interaction energy with the rest of the polarons in the initial and the final positions, using the relative permittivity $\epsilon_r = 20$. The interface between the sample and the vacuum was resolved analytically by using the mirror charges $q'=(\epsilon-1)/(\epsilon+1)q$, where $q$ is the charge of one polaron.[41,42] The simulation includes the electrostatic interaction between all polarons, leading to a preferential movement in the direction against the gradient of the potential, resulting in spreading of the polaronic cloud. The potential above the surface corresponding to the LCPD was calculated using the effective value of the dielectric permittivity $\epsilon_{eff}=(\epsilon_r+1)/2$.[41] Further details are provided in the SI.

Spin-resolved density-functional theory (DFT) calculations were performed using the Vienna *ab initio* software package (VASP).[43,44] We adopted the HSE06 hybrid functional[45] with 12% of Hartree Fock mixing,[28] a plane-wave energy cutoff of 550 eV, and the Γ point for sampling the reciprocal space. The hematite unitcell was built using a 4×4×1 super cell (with 480 atoms). The Ti doping was modeled by substituting one Fe atom (corresponding to ~0.5% doping). In the undoped cells, the excess hole/electron required to form a polaron was introduced by modifying the total number of electrons and adding a compensating background charge to keep overall charge neutrality. All the atomic coordinates were relaxed to obtain residual forces smaller than 0.01 eV/Å. The polaronic hopping was modelled by linearly interpolating the atomic coordinates between the initial and the final structures. The charge-density isosurfaces were calculated for the in-gap polaronic peaks appearing in the density of states (DOS). The degree of localization of polarons was evaluated by a projection of the polaronic state into the local atomic orbitals (the corresponding local magnetic moments match the values from previous analysis). [28]

Acknowledgments:
The work was funded by projects GACR20-21727X, GAUK Primus/20/SCI/009 (JRed, DW, AA and MSe) and by the Austrian Science Fund FWF (Special Research Program, F8100-N, TACO). Support from the joint Austrian (BMBWF CZ15/2021) and Czech (MSMT8J21AT004) project (by MR, MSe and FE) is acknowledged. DW acknowledges the Bekker NAWA stipend PPN/BEK/2020/1/00415. GSP and FK acknowledge funding from the European Research Council (ERC) under the European Union's Horizon 2020 research and innovation programme (grant agreement No. [864628], Consolidator Research Grant 'E-SAC').